# A brief introduction to quantum algorithms

Shihao Zhang[1]    Lvzhou Li[1, †]

1 Institute of Quantum Computing and Computer Theory, School of Computer Science and Engineering, Sun Yat-Sen University, Guangzhou 510006, China

† Corresponding author: lilvzh@mail.sysu.edu.cn

## Abstract

Quantum algorithms are demonstrated to outperform classical algorithms for certain problems and thus are promising candidates for efficient information processing. Herein we aim to provide a brief and popular introduction to quantum algorithms for both the academic community and the general public with interest. We start from elucidating quantum parallelism, the basic framework of quantum algorithms and the difficulty of quantum algorithm design. Then we mainly focus on a historical overview of progress in quantum algorithm research over the past three to four decades. Finally, we clarify two common questions about the study of quantum algorithms, hoping to stimulate readers for further exploration.

## 1. Introduction

In recent years, the field of quantum computing has been attracting a great deal of attention (Nielsen and Chuang 2010; Preskill 2018; Gyongyosi and Imre 2019). The fundamental reason is that the intrinsic powerful parallelism of this computing paradigm enables it to effectively solve some problems that are hard for classical computers. However, the parallelism of quantum computing is not directly available, but rather plays a role only through clever algorithm design according to the problem to be solved (Deutsch 1985; Chuang and Yamamoto 1995; Galindo and Martin-Delgado 2002; Paredes, Verstraete et al. 2005; Bravyi, Gosset et al. 2018; Watts, Kothari et al. 2019). Even if one day quantum computer hardware is developed into a mature enough stage, the potential of quantum computing may still not be substantively achieved without appropriate quantum algorithms. Therefore, the study of quantum computing needs to focus on both software and hardware aspects.

With the development of quantum algorithms over more than past thirty years, there have been comprehensive and informative review articles written at different stages for readers with more or less knowledge about quantum computing and quantum information processing (Cleve, Ekert et al. 1998; Shor 2004; Mosca 2008; Bacon and van Dam 2010; Childs and van Dam 2010; Montanaro 2016). As a complement, this survey article aims to provide a more popular introduction to a selection of significant subjects about quantum algorithms for scholars as well as the general public with interest, and is organized as follows. In Sect. 2 we explore parallelism as a fundamental feature in quantum computing from the perspective of quantum physics. In Sect. 3, we outline the basic framework of quantum algorithms, followed by illustrating the difficulty in designing quantum algorithms in Sect. 4. In Sect. 5 we present a historical overview of progress in quantum algorithm research, hoping to stimulate readers' interest in further exploration. Also, we answer two common questions about quantum algorithms for beginners in Sect. 6 by pointing out the significance and methods in this area. Sect. 7 concludes this paper.

## 2 Parallelism of quantum computing

What is the root of quantum parallelism in quantum computing? Here we give the following comments:

**(1) Quantum superposition leads to potential parallelism.** The so-called quantum superposition refers to that a quantum system can exist in a linear combination of different basis states as

$$|\psi\rangle = \alpha_{00...0}|00...0\rangle + \alpha_{00...1}|00...1\rangle + ... + \alpha_{11...1}|11...1\rangle, \quad (1)$$

where the coefficients $\alpha_{00...0}, \alpha_{00...1}, ..., \alpha_{11...1}$ are called the *probability amplitudes* and can take complex values. For a quantum bit (qubit) in the superposition state

$$|\psi_{in}\rangle = \alpha|0\rangle + \beta|1\rangle, \quad (2)$$

one unitary operation $U_f$ can act on its two basis states $|0\rangle$ and $|1\rangle$ in parallel and thus obtain the output state

$$|\psi_{out}\rangle = U_f|\psi_{in}\rangle = \alpha U_f|0\rangle + \beta U_f|1\rangle. \quad (3)$$

This procedure is recognized as "one operation performs two calculations at the same time", and can be generalized to deal with the cases of multi-qubit quantum states. However, note that this parallelism is *not* immediately useful for solving a certain problem, since what we can get is a superposition state and the follow-up elaborate algorithm design is needed to extract the desired information. As Aaronson said (Aaronson 2013), "From reading newspapers, magazines, and so on, one would think a quantum computer could 'solve NP-complete problems in a heartbeat' by 'trying every possible solution in parallel', and then instantly picking the correct one. Well, arguably that's the central misconception about quantum computing among laypeople."

**(2) Interference makes it possible to take advantage of parallelism.** In the middle school physics class, everyone has learned a physical phenomenon called "double-slit interference", that is, the light waves from a coherent light source passing through a plate pierced by two parallel slits can interfere and thus produce bright and dark bands on the screen behind the plate. From a mathematical point of view, interference can be roughly understood as when several paths with different weights (which can take complex numbers) converge, these weights may add up to a large value (i.e. constructive interference) or in a way that cancel each other out (i.e. destructive interference). An example of interference is depicted in Fig. 1.

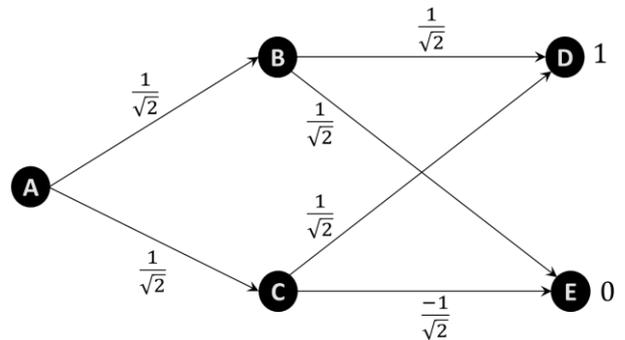

**Fig. 1 Schematic diagram of interference.**

As shown in Fig. 1, there are two paths from the start point A to the end point D, where the path $A \to B \to D$ with the weight $1/\sqrt{2} \times 1/\sqrt{2} = 1/2$ and the path $A \to C \to D$ with the weight $1/\sqrt{2} \times 1/\sqrt{2} = 1/2$ add up to the value 1. As comparison, the path $A \to B \to E$ with the weight $1/\sqrt{2} \times 1/\sqrt{2} = 1/2$ and the path $A \to C \to E$ with the weight $1/\sqrt{2} \times (-1/\sqrt{2}) = -1/2$ from the start point A to the end point E would cancel each other out to the value 0.

For general quantum programs, whether we can use this interference property to make the parallelism of quantum computing available to us requires a very intelligent algorithm design. The key is to exploit interference to increase the weights of the target paths we want, but to decrease those of the undesired paths as much as possible. For readers to comprehend such quantum interference effects in more depth, we'd like to suggest a bunch of relevant quantum-walk-based algorithms as candidates for further study (Childs, Cleve et al. 2003; Ambainis 2007; Schreiber, Cassemiro et al. 2010; Qiang, Wang et al. 2021).

In summary, the potential parallelism rooted in quantum superposition is likely to be utilized by the interference effect. Therefore, the designed algorithms exploiting quantum superposition and interference can take advantage of parallelism for tackling practical problems.

## 3 The basic framework of quantum algorithms

The key to the design of a quantum algorithm is to ensure that each step of the algorithm meets the requirements of quantum mechanics, which ultimately can solve the target problem faster than the classical algorithm in the computational complexity sense (Bacon and van Dam 2010; Montanaro 2016). The problem that can be solved quickly by taking advantage of the parallelism of quantum computing is usually a certain global property denoted $P(f)$ of the function $f$ in mathematics, which depends on function values at multiple points in a certain interval such as the period of $f$. In order to give readers an intuitive impression, we present the framework that consists of three basic steps of many quantum algorithms (Svore 2016; Abhijith 2020) :

**(1) To prepare a quantum superposition state**

$$\alpha_1|x_1\rangle + \alpha_2|x_2\rangle + \cdots + \alpha_N|x_N\rangle \qquad (4)$$

with normalized complex amplitudes $\sum_{i=1}^{N}|\alpha_i|^2 = 1$, which represents the linear combination of the independent variable values $X = \{x_1, x_2, \text{K}, x_N\}$ of the function $f$ ;

**(2) To apply the linear operator (matrix)** $U_f$ **associated with the function $f$ to the superposition state in stage (1).** The linearity of $U_f$ enables it to act on each basis state of Eq. (4) separately to evaluate the function $f(x_i)$ for each independent variable $x_i \in X$ simultaneously and obtain the state

$$\alpha_1|f(x_1)\rangle + \alpha_2|f(x_2)\rangle + \cdots + \alpha_N|f(x_N)\rangle, \qquad (5)$$

which embodies the potential parallelism;

**(3) To exact desired information.** By clever design, the use of interference onto Eq. (5) makes the final state of the system fall into the target state $|P(f)\rangle$ with great probability, and the desired information can be obtained by measurement. The ingenuity of the algorithm design is reflected in this step.

Here we exemplify this procedure for the two-qubit case of the well-known Grover's search algorithm (Grover 1996, 1997; Walther et al. 2005; Diao 2010) for elaboration, which can be represented by a oracle function $f(x)$ defined by

$$f(x)=\begin{cases}1, & x=x_0 \\ 0, & x \neq x_0\end{cases} \quad (6)$$

with $x_0$ being the target solution. Then the procedure is as follows: (i) to prepare an initial two-qubit equal superposition state

$$|\psi_{ini}\rangle = \frac{1}{\sqrt{2}}(|00\rangle+|01\rangle+|10\rangle+|11\rangle); \quad (7)$$

(ii) to apply the oracle operator $U_f: |x\rangle \rightarrow (-1)^{f(x)}|x\rangle$ to the state $|\psi_{ini}\rangle$ of Eq. (7) and obtain

$$|\psi_f\rangle = \frac{1}{\sqrt{2}}[(-1)^{f(00)}|00\rangle+(-1)^{f(01)}|01\rangle \\ +(-1)^{f(10)}|10\rangle+(-1)^{f(11)}|11\rangle] \quad (8)$$

which marks the solution state $|x_0\rangle$ by shifting its phase; (iii) to apply the inversion about mean operator $(2|\psi_{ini}\rangle\langle\psi_{ini}|-I)$ to $|\psi_f\rangle$ of Eq. (8) with $I$ being the identity operator and it can be verified the output state is exactly $|x_0\rangle$. Thus, the target solution $x_0$ is obtained by measuring this output state in computational bases.

Note the framework introduced above is basic yet probably not all-powerful for quantum algorithm design, and the actually employed approaches would be more complicated and flexible according to the properties of specific problems (Mosca 2008; Childs and van Dam 2010; Shao, Li et al. 2019). In the following we analyze the difficulty of putting forward excellent quantum algorithms.

## 4 The difficulty of quantum algorithm design

During the line of research on quantum computing, experts have realized that the design of good quantum algorithms is quite challenging (Shor 2003; Shor 2004). The difficulty is mainly reflected in the following two aspects:

(1) To come up with thoughtful algorithms is always challenging, even in the field of classical computing. Any original algorithm is the crystallization of wisdom.

(2) The counter-intuitive nature of quantum mechanics, that is, strange features of quantum mechanics compared to classical physics as indicated in Section 2.2.9 of the textbook by (Nielsen and Chuang 2010), might make the algorithm design experience accumulated in the classical world no longer applicable. For example, a classical CNOT gate can fulfill the task of copying a classical bit, whereas the *no-cloning theorem* tells us it is impossible to make an ideal copy of an unknown quantum state (Wootters and Zurek 1982). Another typical example of a fundamental difference between quantum and classical physics is the *violation of Bell's inequality* by performing specific measurements on the entangled quantum state (Bell 1964; Clauser et al. 1969), which contradicts the assumptions of *local realism* from the classical view. Considering these intrinsic differences, the task to design a quantum algorithm better than the classical algorithm makes it even more difficult. At present, people are not quite clear about the scope of the computational problems that can reflect quantum advantage, and basically cross the river by feeling the stones for designing efficient quantum algorithms.

For these reasons, how to overcome the difficulties and build novel quantum algorithms that can offer a substantial computational speed-up over classical counterparts for solving important problems is a long-term goal. Nowadays, the newly proposed methodologies have derived inspiration from the emerging field of machine learning and artificial intelligence and made noteworthy progress (Bang, Ryu et al. 2014; Dunjko and Briegel 2018; Lin, Lai et al. 2020).

## 5 Overview of progress in quantum algorithm research

Although it is not easy to design remarkable quantum algorithms, researchers have made a lot of efforts and achieved a series of progress. There is often a saying that "there are just several quantum algorithms now", which is true to some extent since there are indeed not so many well-known representative quantum algorithms in public perception. But from another perspective, the above statement is imprecise due to the existence of hundreds of quantum algorithms for different application scenarios. Readers can refer to the website https://quantumalgorithmzoo.org/ for learning more about the current major quantum algorithms.

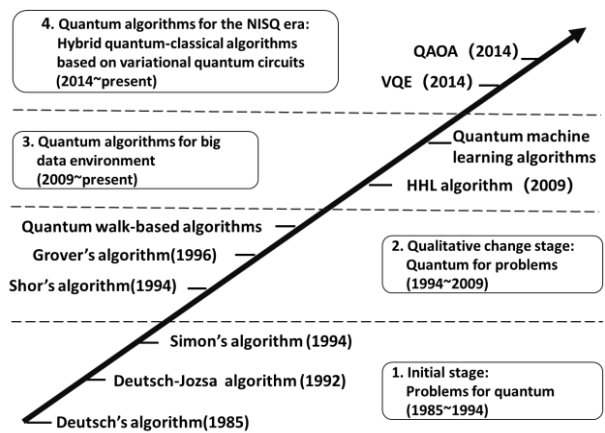

**Fig. 2 A historical overview of progress in quantum algorithm research.**

As the timeline in Fig. 2 shows, the history of quantum algorithms can be roughly divided into four stages:

**The first stage of quantum algorithms (1985-1994), which we call the initial stage, is characterized by "problems for quantum".** That is, some mathematical problems are constructed and quantum algorithms are designed for them in order to demonstrate the advantages of quantum computing, which may not be of much practical value at that time. The earliest quantum algorithm can be traced back to Deutsch's algorithm (Deutsch 1985). In 1985, David Deutsch considered a simple question in his seminal paper about quantum Turing machines, and designed a quantum computing process for it. By utilizing quantum superposition and interference, this constructed example suggested that quantum computers might have computational powers exceeding those of classical computers and planted the seeds of ideas for subsequent quantum algorithm design. Although it seems that Deutsch algorithm is very simple today and may even be taken for granted, the design of the first prototype of the quantum algorithm in that unprecedented era required insight and creativity indeed. The subsequent Deutsch-Jozsa algorithm (Deutsch and Jozsa 1992) and Simon's algorithm (Simon 1994) further considered more complex problems and in a sense demonstrated the advantage of quantum computing as an exponential speedup over classical computing.

By the way, Simon's algorithm may be a somewhat neglected quantum algorithm. In fact, the significance of this algorithm includes at least the following two aspects:

(1) Simon's algorithm directly inspired the discovery of the famous Shor's algorithm, which is clearly pointed out in Shor's original paper (Shor 1994) or in books on quantum computing (Gruska 1999; Nielsen and Chuang 2010; Williams 2011).

(2) Simon's algorithm has been applied to the field of cryptography in recent years. Although the problem it solves did not exhibit obvious application scenarios at the beginning of its appearance, the research into breaking cryptosystems based on Simon's algorithm has been followed up in later years (Boneh and Lipton 1995; Okamoto, Tanaka et al. 2000; Kaplan, Leurent et al. 2016).

**The second stage of quantum algorithms (1994-2009), which we call the qualitative change stage, is characterized by "quantum for problems"**, which aims to design ingenious quantum algorithms for specific

issues of important application values. In 1994, Shor's algorithm (Shor 1994) showed that the factoring problem can be solved by quantum computers in polynomial time, while the difficulty of this problem for classical computers is the theoretical basis of the security of the RSA public-key cryptosystem. Subsequently, in 1996 Grover (Grover 1996) discovered a quantum algorithm for searching unstructured databases that can offer a quadratic speedup, making it possible to "search for a needle in a haystack" more quickly than best possible classical algorithms (Grover 1997). Since the problems solved by these algorithms have a wide range of application values, they have attracted much attention and greatly promoted the development of the entire quantum computing field. A lot of follow-up studies focused on how to adapt the above two algorithms to more practical problems. In addition, the quantum walk (Ambainis 2003; Childs, Cleve et al. 2003; Ambainis 2007) proposed at this stage also provided a useful tool for quantum algorithm design.

**The third stage (2009-present) of quantum algorithms, which we call the new development stage, is characterized by facing with big data environment.** The introduction of the Harrow-Hassidim-Lloyd (HHL) algorithm (Harrow, Hassidim et al. 2009) for solving linear equations in 2009 marked the third stage of quantum algorithms. Perhaps HHL algorithm is not comparable to Shor's algorithm or Grover's algorithm, but it opened up a new avenue for the design of quantum algorithms after everyone has been waiting for the emergence of new quantum algorithms for more than 10 years. This may bring into a new paradigm of applying quantum simulation to data processing. Quantum simulation is an important aspect of quantum computing, which also involves the research in various simulation algorithms. However, since this paper mainly focuses on the use of quantum information technology for classical data processing, the topic of quantum simulation more related to quantum physical processes will not be introduced in detail here.

**Table 1 Typical Quantum machine learning algorithms.**

| | |
|---|---|
| Regression | Quantum linear regression (Wiebe, Braun et al. 2012; Wang 2017); Quantum ridge regression (Liu and Zhang 2017; Yu et al. 2019, 2021) |
| Classification | Quantum SVM/SMM (Rebentrost, Mohseni et al. 2014; Duan et al. 2017; Ye et al. 2020) |
| | Quantum kNN (Wiebe, Kapoor et al. 2015) |
| Dimensionality reduction | Quantum PCA/LDA (Lloyd, Mohseni et al. 2014; Cong and Duan 2016); Quantum AOP (Duan et al. 2019; Pan et al. 2020) |
| Neural networks | Quantum perceptron (Kapoor, Wiebe et al. 2016; Ban et al. 2021) |
| | Quantum deep learning (Wiebe, Kapoor et al. 2016; Li et al. 2020; Shen et al. 2020; Yang and Zhang 2020) |
| Recommendation System (RS) | Quantum RS (Kerenidis and Prakash 2017) |
| | Quantum-inspired RS algorithm (Tang 2019; Du et al. 2020) |
| Learning theory | Quantum learning theory (Arunachalam and De Wolf 2017) |
| Generative adversarial network (GAN) | Quantum GAN (Dallaire-Demers and Killoran 2018; Lloyd and Weedbrook 2018; Situ, He et al. |

| | 2020; Zeng et al. 2019; Hu et al. 2019) |

Since numerous methods and technologies in the field of artificial intelligence (AI) and big data are related to solving linear equations, HHL algorithm has greatly promoted quantum computing into the fields of machine learning and big data processing. The combination of quantum computing and AI has become a hot topic in recent years. Several typical achievements related to quantum machine learning are listed in Table 1, and there is no doubt the cross-over studies in this area are worthy of more in-depth exploration (Schuld, Sinayskiy et al. 2015; Yu et al. 2016; Biamonte, Wittek et al. 2017; Dunjko and Briegel 2018; Wang et al. 2018; Allcock and Zhang 2019). Also, here we make a few comments on quantum machine learning algorithms:

(1) The HHL algorithm does not present the solution of the equation system in a classically readable manner, but encodes it in a quantum state that requires subsequent algorithm design to extract the information we want. A large number of studies on quantum machine learning in recent years are mainly based on the HHL algorithm for subsequent algorithm design.

(2) At present, more serious theoretical analyses need to be provided for some research on quantum machine learning besides accessible numerical simulation.

(3) When quantum machine learning faces the actual data processing problem, it needs to break through the input / output bottleneck. Here the so-called input / output bottleneck means that most of the current quantum machine learning algorithms either need to encode large-scale data sets into quantum states, or just generate the solution to the problem in the quantum state. In this setting, the pre-processing in the input stage and the post-processing in the information extraction stage will consume a lot of time, which even offsets the time saved by the quantum algorithm.

Recently, Tang has designed a classical algorithm inspired by quantum recommendation algorithm (Tang 2019), which can solve the recommendation problem only polynomially slower than that carried by quantum algorithm. In this way, his algorithm produces recommendations exponentially faster than previous classical systems under certain input assumptions, and the research direction of quantum-inspired classical algorithm design or "dequantizing" quantum algorithms has attracted more scholars' attention so far (Arrazola, Delgado et al. 2020; Chia, Gilyén et al. 2020; Deng et al. 2020; Jethwani, Le Gall et al. 2020; Ding, Bao et al. 2021). If the thought of quantum algorithms can facilitate the development of classical algorithms, it will also be another embodiment of the significance of quantum computing research.

**The fourth stage (2014-present) of quantum algorithm is characterized by quantum algorithms for the noisy intermediate-scale quantum (NISQ) era (Preskill 2018).** In recent years the so called "quantum supremacy" or "quantum advantage" has been achieved experimentally in superconducting circuits (Arute, Arya et al. 2019) and photonic systems (Zhong, Wang et al. 2020) at certain sampling tasks, and exploring more useful computational tasks and dedicated algorithms well-suited for implementation on NISQ devices is of great interest. Hybrid quantum-classical algorithms arisen at this stage, including the representative variational quantum eigensolvers (VQE) (Peruzzo, McClean et al. 2014) and quantum approximate optimization algorithms (QAOA) (Farhi, Goldstone et al. 2014), are rapidly growing and expected to promote the availability of quantum computing (McClean, Romero et al. 2016; Endo, Cai et al.

2021). By cooperating quantum computers and classical computers, the designed hybrid quantum-classical approaches have brought a wide range of applications with both theoretical progress and experimental demonstrations, such as aiding quantum chemistry simulation (Peruzzo, McClean et al. 2014; Kandala, Mezzacapo et al. 2017; Moll, Barkoutsos et al. 2018), solving general machine learning problems (Mitarai, Negoro et al. 2018; Benedetti, Lloyd et al. 2019; Jones, Endo et al. 2019; Shao 2019; Liang et al. 2020; Schuld, Sweke et al. 2021; Wang, Song et al. 2021; Xu et al. 2021), promoting quantum error mitigation methods (Li and Benjamin 2017; McArdle, Yuan et al. 2019; Otten and Gray 2019; Huggins, McClean et al. 2021) and quantum materials science (Bauer, Bravyi et al. 2020).

## 6 Two questions about quantum algorithms

Here we attempt to answer two common questions about the study of quantum algorithms.

**Question 1:** Is it necessary to study quantum algorithms before the appearance of practically useful quantum computers?

We answer the above question as follows:

(1) From the perspective of classical computing, the research on algorithms predates the emergence of practical computers. Euclid's algorithm appeared in ancient Greece, while the first electronic computer was produced in the year 1946. In addition, one of the purposes of proposing Turing machine was to strictly describe "algorithm" (Turing 1937).

(2) Research on general or specific quantum algorithms is also one of the important parts in the development of quantum computers. Quantum algorithms are unarguably the necessary software support for quantum computers, and the research of quantum algorithms is also a powerful driving force to promote the progress of quantum computing, which can be seen from the historical status of Shor's algorithm.

**Question 2:** How to study quantum algorithms and evaluate their performances without realistic quantum computers?

We answer the above question as follows:

(1) Abstract-level algorithm design doesn't have to depend on a specific hardware platform, which is like the case in the field of classical computing. Algorithms are essentially methods invented for solving particular problems, and quantum algorithms are methods following the laws of quantum mechanics. At the hardware level, the built quantum platform is just a tool to actually demonstrate this quantum algorithm. Of course, the interaction and communication between software and hardware would be beneficial to the design of more realistic algorithms, e.g., by providing parameters and data from practical experiments.

(2) From the perspective of algorithm and complexity research, the quality of an algorithm is measured by computational complexity that relies on strict mathematical proofs (except for some heuristic algorithms) rather than testing on specific hardware platforms. At present, the mainstream research on quantum algorithms is like this.

(3) With the development of experimental quantum computing technology, more and more quantum algorithms can be verified in principle on the real quantum platforms (Figgatt, Maslov et al. 2017; Linke, Maslov et al. 2017; Zhong, Wang et al. 2020). Sometimes researchers may simulate or emulate quantum algorithms on classical computing devices to deepen our understanding (Bravyi and Gosset 2016, Chen, Zhou et al. 2018, Guo, Liu et al. 2019, Zhang, Zhang et al. 2019, Kobori, Takahashi et al. 2020), but that does not mean the actual realization of quantum algorithms by classical computers.

# 7 Summary

In what aspects do quantum computing have advantages over classical computing? To what extent do relevant advantages retain? These issues are far from clear at present, indicating that there is great potential for the study of quantum algorithms. Everyone is looking forward to the emergence of more innovative algorithms in the field of quantum computing, and correspondingly every quantum algorithm researcher also hopes to design a representative algorithm.

However, as the old saying goes: "Rome was not built in a day" or "a journey of thousand miles begins with a single step". We should not just remember the cleverness of Shor's algorithm, but forget the efforts of predecessors. In fact, Shor's algorithm is on the shoulder of Simon's algorithm, which originates from the superficially useless Deutsch-Jozsa and Deutsch's algorithms as well.

This process just reflects the charm of scientific research: perhaps many research results will be washed away by the waves of time, but it is every little bit of or even seemingly unnecessary research that gives birth to a new discovery step by step.